\documentclass{iaus}
\usepackage{graphicx}
\usepackage{natbib}

\newcommand{\aap}{{A{\&}A}}
\newcommand{\aapr}{{A{\&}ARv}}
\newcommand{\apj}{{ApJ}}
\newcommand{\apjl}{{ApJL}}
\newcommand{\nat}{{Nature}}
\newcommand{\pasj}{{Publ. Astron. Soc. Jpn.}}

\title[Numerical sunspot models] 
{3D numerical MHD modeling of sunspots with radiation transport}

\author[Matthias Rempel]   %% give here short author list %%
{Matthias Rempel}

\affiliation{High Altitude Observatory, National Center for Atmospheric 
  Research, \\ P.O. Box 3000, Boulder, CO 80307, USA 
  \\ email: {\tt rempel@ucar.edu}}
 
\pubyear{2010}
\volume{273}  %% insert here IAU Symposium No.
\pagerange{1-6}
\jname{Physics of Sun and Star Spots}
\editors{}

\begin{document}

\maketitle

\begin{abstract}
Sunspot fine structure has been modeled in the past by a combination
of idealized magneto-convection simulations and simplified models that
prescribe the magnetic field and flow structure to a large degree.
Advancement in numerical methods and computing power has enabled recently
3D radiative MHD simulations of entire sunspots with sufficient resolution
to address details of umbral dots and penumbral filaments. After a brief
review of recent developments we focus on the magneto-convective processes
responsible for the complicated magnetic structure of the penumbra and the
mechanisms leading to the driving of strong horizontal outflows in the
penumbra (Evershed effect). The bulk of energy and mass is transported on
scales smaller than the radial extent of the penumbra. Strong horizontal 
outflows in the sunspot penumbra result from a redistribution of kinetic energy
preferring flows along the filaments. This redistribution is facilitated
primarily through the Lorentz force, while horizontal pressure gradients play 
only a minor role. The Evershed flow is strongly magnetized: While we see a 
strong
reduction of the vertical field, the horizontal field component is enhanced 
within filaments.
\keywords{MHD -- radiative transfer -- Sun: sunspots -- Sun: photosphere 
  -- Sun: magnetic fields}
\end{abstract}

\section{Introduction}
\label{intro}
Sun and starspots play a central role for our understanding of solar and 
stellar magnetism. Since direct observations of magnetic field in stellar
convection zones are very limited (e.g. helioseismic inversions), sun and 
starspots provide a window to understand the magnetism of stellar interiors,
provided we understand in detail how starspots form, evolve and decay.
Sunspots are a multi scale problem with respect to spatial and temporal
scales. While their typical size is $>20$~Mm, fine structure is observed at
the smallest scales currently observable of about $200$~km (e.g. high resolution
ground based observations with the {\it SST}, \citet{Scharmer:etal:2002}, or 
space based observations with {\it HINODE}, see e.g. 
\cite{Ichimoto:etal:2007}). Properly resolving the scales currently
observed requires grid resolutions of $20$~km or less. Details of the penumbra
evolve over time scales of hours, while the life time of sunspots and the
evolution of the adjacent moat happens on time scales of days to weeks. On
the other hand typical numerical time steps are of the order of $0.1$~sec
(assuming already that the very fast Alfv{\'e}n velocities found above a sunspot
umbra were removed from the system, see e.g. appendix of 
\cite{Rempel:etal:2009}). As a consequence a well resolved realistic numerical 
simulation of an entire sunspot requires billions of grid points and 
hundred thousands of time steps. Since the essential physics that need to
be considered (MHD, 3D radiative transfer, realistic equation of state)
are well known and have been included in MHD codes for more than 2 decades,
the primary challenge for addressing sunspot structure is linked to robust
efficient numerical schemes and availability of powerful computing resources. 
The latter became finally available on the scale needed, which allowed for
substantial progress within a time frame of only a few years.
In the following sections we will very briefly summarize recent progress and 
a few aspects of sunspot structure for which numerical simulations provided 
substantial insight.   

\begin{figure}[b]
\begin{center}
 \includegraphics[width=13cm]{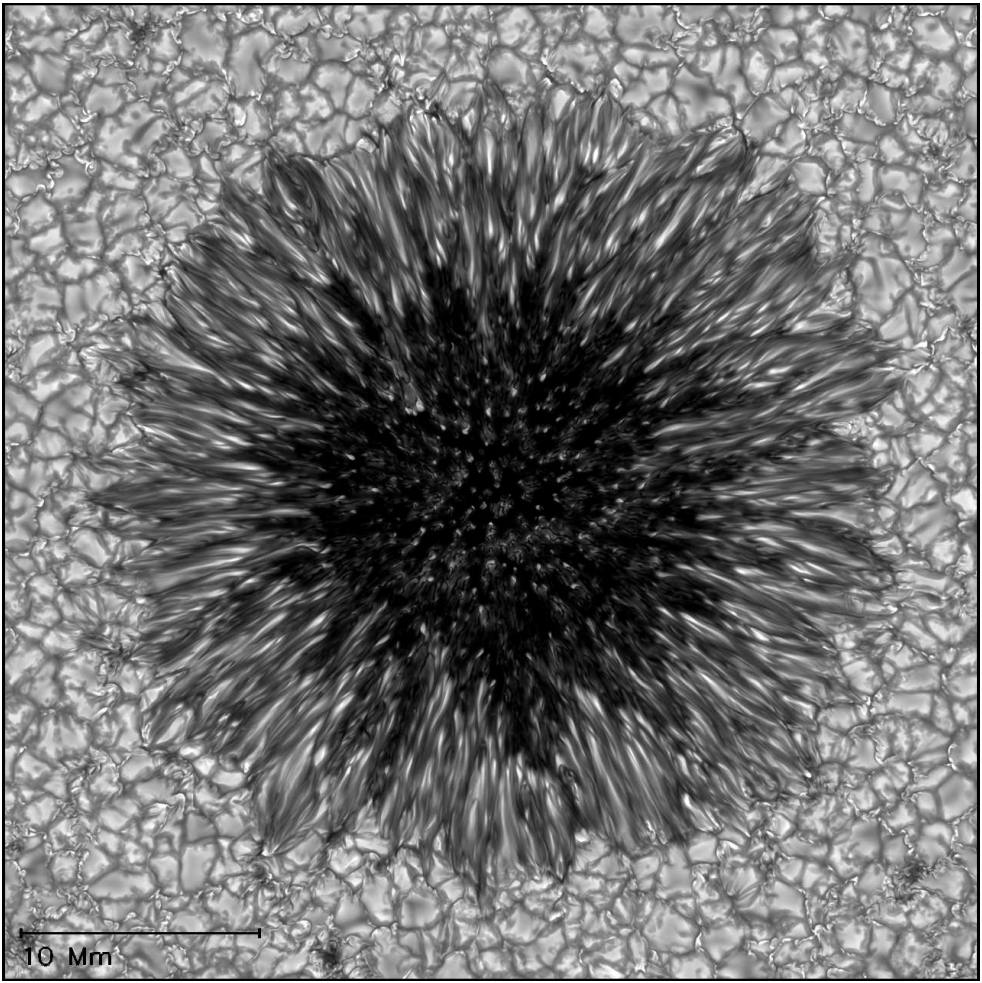} 
% \vspace*{-1.0 cm}
 \caption{Numerical sunspot model. The domain size is $49\times 49\times 6.1$ 
   Mm, the grid resolution $16\times16\times12$ km ($3072\times 3072\times 512$
   grid points). The simulation was performed with grey radiative transfer, 
   displayed is the bolometric intensity for a vertical ray in the range from 
   $0.25 - 1.5 I_{\odot}$.}
  \label{fig1}
\end{center}
\end{figure}

\section{Recent developments}
\label{recent}
We limit the following discussion entirely to MHD simulations that include 
3D radiative transfer and a realistic equation of state. Here, progress in
applications related to sunspots started with the work by 
\cite{Schuessler:Voegler:2006} who presented a MHD simulation of a sunspot 
umbra showing the development of magneto-convection in form of umbral dots.
The simulation revealed that almost field free upflow plumes can form within
an initially monolithic umbra and transport energy through overturning
convection. \cite{Heinemann:etal:2007} focused on a narrow slab through the
center of sunspot. Their setup allowed to model the transition from umbra
toward penumbra and granulation, while keeping the computational expense at a 
moderate level. Their simulation showed the formation of short filaments with 
dark lanes at the umbra/granulation interface, however, umbral dots were not
present. The filaments were propagating inward during the formation phase
and showed outflow along their axis. Based on this simulation 
\cite{Scharmer:etal:2008} interpreted the Evershed flow as convective flow
component along filaments. In a very similar setup with an initially larger
segment of a sunspot  \cite{Rempel:etal:2009} were able to produce umbral dots
and filaments of about $3$ Mm length, showing the smooth transition from 
central to peripheral umbral dots and filaments in the inner penumbra. The 
latter show the presence of a bright head propagating 
inward and detaching from the filament during the formation phase and up to
$3$ Mm long dark lanes with evidence of twisting motions. Along filaments, 
outflows of a few km/s were present, and on a larger scale a moat flow, 
previously
also reported in \cite{Heinemann:etal:2007}. Overall these simulations clearly 
stress the common magneto-convective origin of umbral dots and penumbral 
filaments, but a realistic outer penumbra with strong outflows was not present.
\cite{Kitiashvili:etal:2009} conducted a study of how the magnetic field 
strength
and inclination angle influence the formation of horizontal flows in 
magneto-convection. Fast flows with mean amplitudes of $1-2$ km/sec were found 
for a
$1.5$ kG field with inclination angle of $85$ degrees. The first comprehensive
simulation of entire sunspots was presented by \cite{Rempel:etal:Science}.
Using a setup including a pair of opposite polarity sunspots in a $98\times49$~Mm
wide and $6.1$~Mm deep domain, this simulation showed the 
formation of extended (up to $10$ Mm wide) penumbrae with horizontal outflows
of up to $6$ km/sec mean- and $14$ km/sec peak flow speeds. At the same time
this simulation contained umbral dots and a substantial moat region,
connecting all the aspects of sunspot fine structure in one comprehensive 
simulation run. While the penumbra contained radially elongated 
convection cells, the intensity image (see Fig. 1 in 
\cite{Rempel:etal:Science}) did not yet show the typical radial structure
of narrow filaments known from observations. The currently best resolved sunspot
simulation is presented in Fig. \ref{fig1}. Compared to 
\cite{Rempel:etal:Science} the horizontal resolution is doubled (from $32$ to 
$16$ km), while the vertical resolution is increased from $16$ to $12$ km,
leading to a total grid size of $3072\times 3072\times 512$ for a domain of
the size $49\times 49 \times 6.1$~Mm. 
Here we focused again on a single spot, but artificially enhanced the field
inclination through the top boundary condition to generate conditions that 
are comparable to the region
in between the opposite polarity spots of \cite{Rempel:etal:Science}. While
the photospheric appearance of the sunspot is substantially improved compared
to previous simulations, we did not find fundamental differences in the
underlying magneto-convection process. 
 
In the following section we highlight 
a few central aspects we learned from the simulations summarized above.

\begin{figure}[b]
\begin{center}
 \includegraphics[width=13cm]{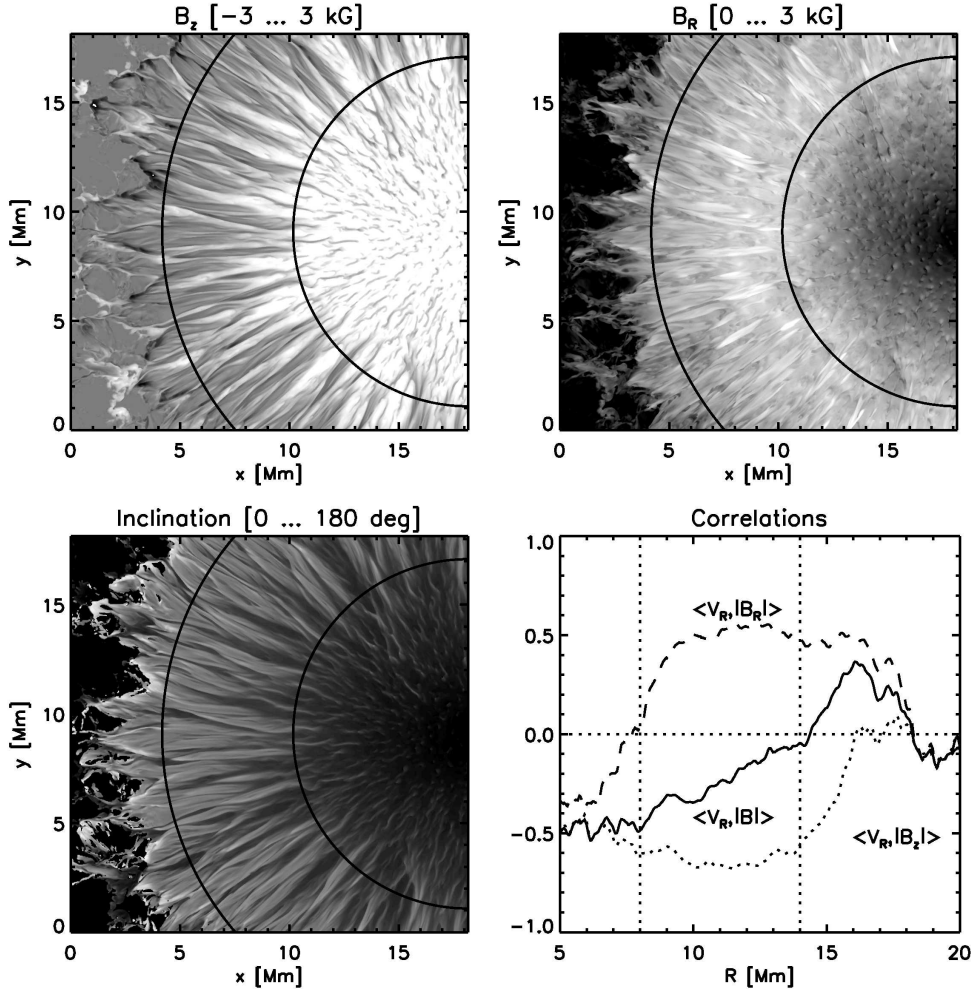} 
% \vspace*{-1.0 cm}
 \caption{Magnetic fine structure of sunspot at $\tau=1$ level. The top panels
   show vertical ($\pm 3$~kG) and radial field strength ($0\ldots 3$~kG). The
   bottom left panel shows the inclination, with black indicating vertical
   outward, grey horizontal, and white vertical inward directed field. We set
   the inclination to $0$ for regions with less than $500$~G field strength
   (dark color outside the penumbra). The bottom right shows correlations 
   between horizontal flow velocity and magnetic field strength fluctuations
   as a function of the spot radius. The vertical dotted lines correspond to 
   the dark concentric circles shown in the other three panels.}
  \label{fig2}
\end{center}
\end{figure}

\section{Magnetic fine structure of penumbra}
Figure \ref{fig2} summarizes the magnetic fine structure of umbra and penumbra
at the $\tau_{\rm Ross}=1$ level for the simulation presented in Fig. \ref{fig1}.
A common element in umbra and penumbra is a strong reduction of the vertical 
magnetic field component, which is associated with umbral dots, peripheral
umbral dots and penumbral filaments. The horizontal magnetic field component
shows a different behavior. A reduction of the horizontal field strength is
only present in the umbra, whilst in the penumbra filaments (flow channels)
have an enhanced horizontal magnetic field strength throughout, which is 
evident from the 
correlations shown in the bottom right panel of Fig. \ref{fig2}. The 
combination of reduced vertical and enhanced horizontal field leads to the
observationally inferred interlocking comb structure with strong variation of
the field inclination angle and strong horizontal flows in the component
with nearly horizontal field (see e.g. \citet{Solanki:2003}) . 
The combined effect of the weakening of
$B_z$ and strengthening of $B_R$ within flow channels results in an overall 
reduction of $\vert B\vert$ in the inner, and increase of $\vert B\vert$ in the
outer, penumbra. Recently \citet{Ichimoto:etal:2007} found this trend in the
observed $V_R-\vert B \vert$ correlation, even though positive values of the
correlation were not observed. An Evershed flow in regions with enhanced field
strength in the outer penumbra was proposed by \cite{Tritschler:etal:2007}
based on zero-crossings of the NCP.

Overall the numerical simulation presented here supports a strongly
magnetized penumbra and Evershed flow. While the underlying structure of the
magneto-convection is well captured by the "gappy" penumbra model of
\cite{Scharmer:Spruit:2006} and \cite{Spruit:Scharmer:2006}, we do not see 
any support for the claim that these gaps are close to field free. 
While the latter is a good approximation for the vertical field component 
alone, horizontal field is actually enhanced compared to the background 
field. Recently \cite{Rempel:2010:submitted}
showed that the horizontal field originates from the strong subsurface shear
profile of the Evershed flow, which leads to an induction term 
$B_z\partial_zv_R$ with a strength of about $5-10$~G/sec a few $100$ km beneath
$\tau=1$. Relative contributions from numerical magnetic diffusivity are only 
of a few percent. A recent convergence study (Rempel, in preparation) from 
$96\times 32$ up to $16\times 12$ km resolution (Fig. \ref{fig1} presents 
the highest resolution case) strongly supports the robustness of this result.

\section{Convective energy transport}
All numerical simulations performed to date (most of them are summarized in
Sect. \ref{recent}) point toward a common magneto-convective origin of
sunspot fine structure. Comparing the azimuthally averaged bolometric 
intensity and vertical RMS velocity at the $\tau_{\rm Ross}=1$ level we find 
a relationship of the form $I\propto\sqrt{v_z^{RMS}(\tau=1)}$, which holds
very well from umbral dots to penumbral filaments and even granulation. 
From this we can conclude that the vertical RMS velocity in the inner penumbra
with $I\approx 0.7\,I_{\odot}$ should be about half of the convective RMS 
velocity seen in granulation, i.e. it should be about $1$ km/sec instead of 
$2$ km/sec. This is, at least in a statistical sense, compatible with the
findings reported in \cite{Franz:Schlichenmaier:2009}, who found about
$500$ m/s vertical RMS in the penumbra vs about $1$ km/sec RMS in the quiet 
sun. The shortfall by a factor of $2$ for both penumbra and quiet sun is
due to the limited resolution of observations and the fact that the typically
used spectral lines form at levels higher than $\tau_{\rm Ross}=1$.

We find throughout the penumbra an upflow filling factor in the $40-60\%$
range and most of the mass flux is turning over on scales significantly
shorter than the radial extent of the penumbra. Only about 
$15\%$ of the total overturning mass flux (similarly also energy flux) is
found in the azimuthal mean component (corresponding to an upflow of about
$200$ m/s in the inner and downflow of up to $500$ m/s in the outer penumbra).
The mass flux of the penumbra is balanced within the bounds of the penumbra.

The almost equal presence of up and downflows everywhere in the penumbra
is currently not inferred from most observations (which see a preference
of upflows in the inner and downflows in the outer penumbra) and might
require even higher resolution. Without overturning motions on scales
much shorter than the radial extent of the penumbra it would not be possible 
to maintain the observed brightness of about $0.7\,I_{\odot}$.

\section{Origin of Evershed flow}
Recently \cite{Rempel:2010:submitted} analyzed in detail the processes
underlying the driving of the Evershed flow. To this end the contributions
from acceleration, pressure, buoyancy and Lorentz forces in the kinetic
energy equation were compared between the plage region (surrounding the 
sunspots) and the sunspot penumbra. In the plage region most of the 
pressure/buoyancy driving takes place in downflows and is in balance with 
vertical acceleration forces. Horizontal flows are driven by horizontal 
pressure gradients. In the penumbra the pressure/buoyancy driving is shifted 
into upflow regions and is there in balance with vertical Lorentz forces, 
while acceleration forces are unimportant. In the horizontal direction
acceleration forces are in balance with the horizontal Lorentz force 
component, while horizontal pressure gradients do not play a major role
throughout most of the penumbra.  
The Lorentz force facilitates the energy exchange
between motions in the vertical and horizontal direction, while the net work 
done by Lorentz forces remains negative (sink of kinetic energy). Some
aspects of this picture (pressure driving in upflows, deflection and guiding 
of motions by magnetic field) have been captured to some extent in simplified 
flux tube models such as \citet{Montesinos:Thomas:1997},
\citet{Schlichenmaier:etal:1998a} and \citet{Schlichenmaier:etal:1998b}. 
However, the presence of vigorous convection leads to a situation in which 
the outflowing mass is continuously replaced through upflows extending along 
filaments -- a situation that is inherently opposite to the concept of a
flux tube. Overall the Evershed flow is best characterized as convective flow 
\citep{Scharmer:etal:2008}, although, compared to field free convection, 
notable differences exist with respect to the underlying driving forces.

\section{Conclusions}
Numerical simulations have been advanced to the point where they provide a 
consistent and unifying picture of the magneto-convective processes
underlying the energy transport, magnetic fine structure and origin of
large scale flows in sunspots. The bulk of energy and mass is transported
by overturning convection with scales substantially shorter than the
radial extent of the penumbra. The filamentation of the penumbra and the
driving of large scale outflows is strongly linked to the presence of
anisotropic convective flows.
We find the above picture being converged with respect to 
numerical resolution. We explored the range from $96\times 32$~km to 
$16\times 12$~km and found that most aspects are already well described with 
$48\times 24$ km resolution. For the 
currently accessible resolution range, flows within filaments are mostly
laminar; whether a possible transition to turbulent flows at higher
resolution could change results remains an open question (due to the rather
strong magnetic field turbulence might remain suppressed even at higher 
resolutions than currently affordable). 
Initial state and boundary conditions at top and
bottom have a strong influence on the global magnetic structure and stability 
(i.e. they can make the difference between having or not having a penumbra),
however the details of sunspot fine structure as discussed here are influenced 
to a much lesser degree. While the overall radial extent of the penumbra is 
subject to boundary conditions, the details of filamentation, energy transport,
and driving mechanism of the Evershed flow are not. 

\acknowledgements
The National Center for Atmospheric Research is sponsored by the National 
Science Foundation. Computing time was provided by the National Center for 
Atmospheric Research, the Texas Advanced Computing Center (TACC), the National 
Institute for Computational Sciences (NICS) and the NASA High End Computing 
Program.

%\bibliographystyle{natbib/apj}
%\bibliography{natbib/papref_m}

\begin{thebibliography}{17}
\expandafter\ifx\csname natexlab\endcsname\relax\def\natexlab#1{#1}\fi

\bibitem[{{Franz} \& {Schlichenmaier}(2009)}]{Franz:Schlichenmaier:2009}
{Franz}, M. \& {Schlichenmaier}, R. 2009, \aap, 508, 1453

\bibitem[{{Heinemann} {et~al.}(2007){Heinemann}, {Nordlund}, {Scharmer}, \&
  {Spruit}}]{Heinemann:etal:2007}
{Heinemann}, T., {Nordlund}, {\AA}., {Scharmer}, G.~B., \& {Spruit}, H.~C.
  2007, \apj, 669, 1390

\bibitem[{{Ichimoto} {et~al.}(2007){Ichimoto}, {Shine}, {Lites}, {Kubo},
  {Shimizu}, {Suematsu}, {Tsuneta}, {Katsukawa}, {Tarbell}, {Title}, {Nagata},
  {Yokoyama}, \& {Shimojo}}]{Ichimoto:etal:2007}
{Ichimoto}, K., {Shine}, R.~A., {Lites}, B., {Kubo}, M., {Shimizu}, T.,
  {Suematsu}, Y., {Tsuneta}, S., {Katsukawa}, Y., {Tarbell}, T.~D., {Title},
  A.~M., {Nagata}, S., {Yokoyama}, T., \& {Shimojo}, M. 2007, \pasj, 59, 593

\bibitem[{{Kitiashvili} {et~al.}(2009){Kitiashvili}, {Kosovichev}, {Wray}, \&
  {Mansour}}]{Kitiashvili:etal:2009}
{Kitiashvili}, I.~N., {Kosovichev}, A.~G., {Wray}, A.~A., \& {Mansour}, N.~N.
  2009, \apjl, 700, L178

\bibitem[{{Montesinos} \& {Thomas}(1997)}]{Montesinos:Thomas:1997}
{Montesinos}, B. \& {Thomas}, J.~H. 1997, \nat, 390, 485

\bibitem[{{Rempel}(2010)}]{Rempel:2010:submitted}
{Rempel}, M. 2010, \apj, submitted

\bibitem[{{Rempel} {et~al.}(2009{\natexlab{a}}){Rempel}, {Sch{\"u}ssler}, \&
  {Kn{\"o}lker}}]{Rempel:etal:2009}
{Rempel}, M., {Sch{\"u}ssler}, M., \& {Kn{\"o}lker}, M. 2009{\natexlab{a}},
  \apj, 691, 640

\bibitem[{{Rempel} {et~al.}(2009{\natexlab{b}}){Rempel}, {Sch{\"u}ssler},
  {Cameron}, \& {Kn{\"o}lker}}]{Rempel:etal:Science}
{Rempel}, M., {Sch{\"u}ssler}, M., {Cameron}, R.~H., \& {Kn{\"o}lker}, M.
  2009{\natexlab{b}}, Science, 325, 171

\bibitem[{{Scharmer} {et~al.}(2002){Scharmer}, {Gudiksen}, {Kiselman}, {L{\"
  o}fdahl}, \& {Rouppe van der Voort}}]{Scharmer:etal:2002}
{Scharmer}, G.~B., {Gudiksen}, B.~V., {Kiselman}, D., {L{\" o}fdahl}, M.~G., \&
  {Rouppe van der Voort}, L.~H.~M. 2002, \nat, 420, 151

\bibitem[{{Scharmer} {et~al.}(2008){Scharmer}, {Nordlund}, \&
  {Heinemann}}]{Scharmer:etal:2008}
{Scharmer}, G.~B., {Nordlund}, {\AA}., \& {Heinemann}, T. 2008, \apjl, 677,
  L149

\bibitem[{{Scharmer} \& {Spruit}(2006)}]{Scharmer:Spruit:2006}
{Scharmer}, G.~B. \& {Spruit}, H.~C. 2006, \aap, 460, 605

\bibitem[{{Schlichenmaier} {et~al.}(1998{\natexlab{a}}){Schlichenmaier},
  {Jahn}, \& {Schmidt}}]{Schlichenmaier:etal:1998b}
{Schlichenmaier}, R., {Jahn}, K., \& {Schmidt}, H.~U. 1998{\natexlab{a}}, \apj,
  493, L121

\bibitem[{{Schlichenmaier} {et~al.}(1998{\natexlab{b}}){Schlichenmaier},
  {Jahn}, \& {Schmidt}}]{Schlichenmaier:etal:1998a}
---. 1998{\natexlab{b}}, \aap, 337, 897

\bibitem[{{Sch{\"u}ssler} \& {V{\"o}gler}(2006)}]{Schuessler:Voegler:2006}
{Sch{\"u}ssler}, M. \& {V{\"o}gler}, A. 2006, \apjl, 641, L73

\bibitem[{{Solanki}(2003)}]{Solanki:2003}
{Solanki}, S.~K. 2003, {\aapr}, 11, 153

\bibitem[{{Spruit} \& {Scharmer}(2006)}]{Spruit:Scharmer:2006}
{Spruit}, H.~C. \& {Scharmer}, G.~B. 2006, \aap, 447, 343

\bibitem[{{Tritschler} {et~al.}(2007){Tritschler}, {M{\"u}ller},
  {Schlichenmaier}, \& {Hagenaar}}]{Tritschler:etal:2007}
{Tritschler}, A., {M{\"u}ller}, D.~A.~N., {Schlichenmaier}, R., \& {Hagenaar},
  H.~J. 2007, \apjl, 671, L85

\end{thebibliography}

\end{document}